\documentclass[aps,prb,twocolumn,amsmath,amssymb]{revtex4}

\usepackage{graphicx}
\usepackage{dcolumn}
\usepackage{bm}
\usepackage[caption=false]{subfig}
\usepackage{amssymb}
\usepackage{amsmath}
\usepackage{graphicx}
\usepackage{verbatim}
\usepackage[usenames, dvipsnames]{color}
\usepackage{hyperref}
\usepackage{color}

\def\vR{{\bf R}}
\def\beq{\begin{equation}}
\def\eeq{\end{equation}}
\def\vk{{\bf k}}

\begin{document}
	
\title{Quadrupolar Superexchange Interactions, Multipolar Order and Magnetic Phase Transition in UO$_2$}
	
	\author{Leonid V. Pourovskii$^{1,2}$ and Sergii Khmelevskyi$^3$}
	
	\affiliation{
		$^1$CPHT, Ecole Polytechnique, CNRS, Universit\'e Paris-Saclay, Route de Saclay, 91128 Palaiseau, France \\
		$^2$Coll\`ege de France, 11 place Marcelin Berthelot, 75005 Paris, France \\
		$^3$Center for Computational Materials Science, IAP, Vienna University of Technology, Vienna, Austria}
\date{\today}
\begin{abstract}
	The origin of non-collinear magnetic order in UO$_{2}$ is studied by an \textit{ab
		initio} dynamical-mean-field-theory framework in conjunction with
	a linear-response approach for evaluating inter-site superexchange
	interactions between U 5$f^{2}$ shells. The calculated quadrupole-quadruple
	superexchange interactions are found to unambiguously resolve the frustration of face-centered-cubic U sublattice
	toward stabilization of the experimentally
	observed non-collinear 3\textbf{k}-magnetic order. Therefore, the
	exotic 3\textbf{k} 
	antiferromagnetic order in UO$_{2}$ 
	 can be 
	 accounted
	for by a purely electronic exchange mechanism acting in the undistorted cubic lattice structure. 
	The quadrupolar short-range order above magnetic ordering temperature $T_N$ is found to
	qualitatively differ from the long-range order below $T_N$. 
\end{abstract}
\maketitle

\section{Introduction}

The interplay of local spin and orbital degrees of freedom (DOF) in strongly
correlated electron systems is at the origin of such remarkable phenomena
as the multiferroic behavior \cite{Spaldin2017}, dynamical single-ion
and cooperative Jahn-Teller effects \cite{kugel_khomskii}, and colossal
magnetoresistance \cite{Tokura}. In rare-earth and actinides compounds
with localized $f$ shells a strong spin-orbit coupling (SOC) in conjunction
with the crystal-field (CF) splitting may lead to emergence of local
multipolar DOF. Intersite interactions between such
multipolar moments in many cases result in their ordering; exotic
multipolar-ordered (MO) states might coexist with the usual magnetic
one \cite{Santini2009}. The rich physics of the multipolar  DOF in $f$-electron systems ranges from the quadrupole interaction
mediated superconductivity\cite{Kotegawa2003} and quadrupolar Kondo
effects\cite{Yatskar} to phonon-mediated electric multipolar interactions
and the dynamical Jan-Teller effect.  Multipolar order parameters are invisible to conventional neutron-diffraction probes
 and thus  notoriously difficult to unambiguously identify experimentally. The quantitative modeling of MO phenomena also represents a significant theoretical challenge  due to a large number of multipolar DOF and a rather small magnitude of  relevant energy scales compared to the conventional Heisenberg dipole-dipole couplings \cite{Caciuffo2011}.  
 
\begin{figure}[!tb]
	\begin{centering}
		\includegraphics[width=0.9\columnwidth]{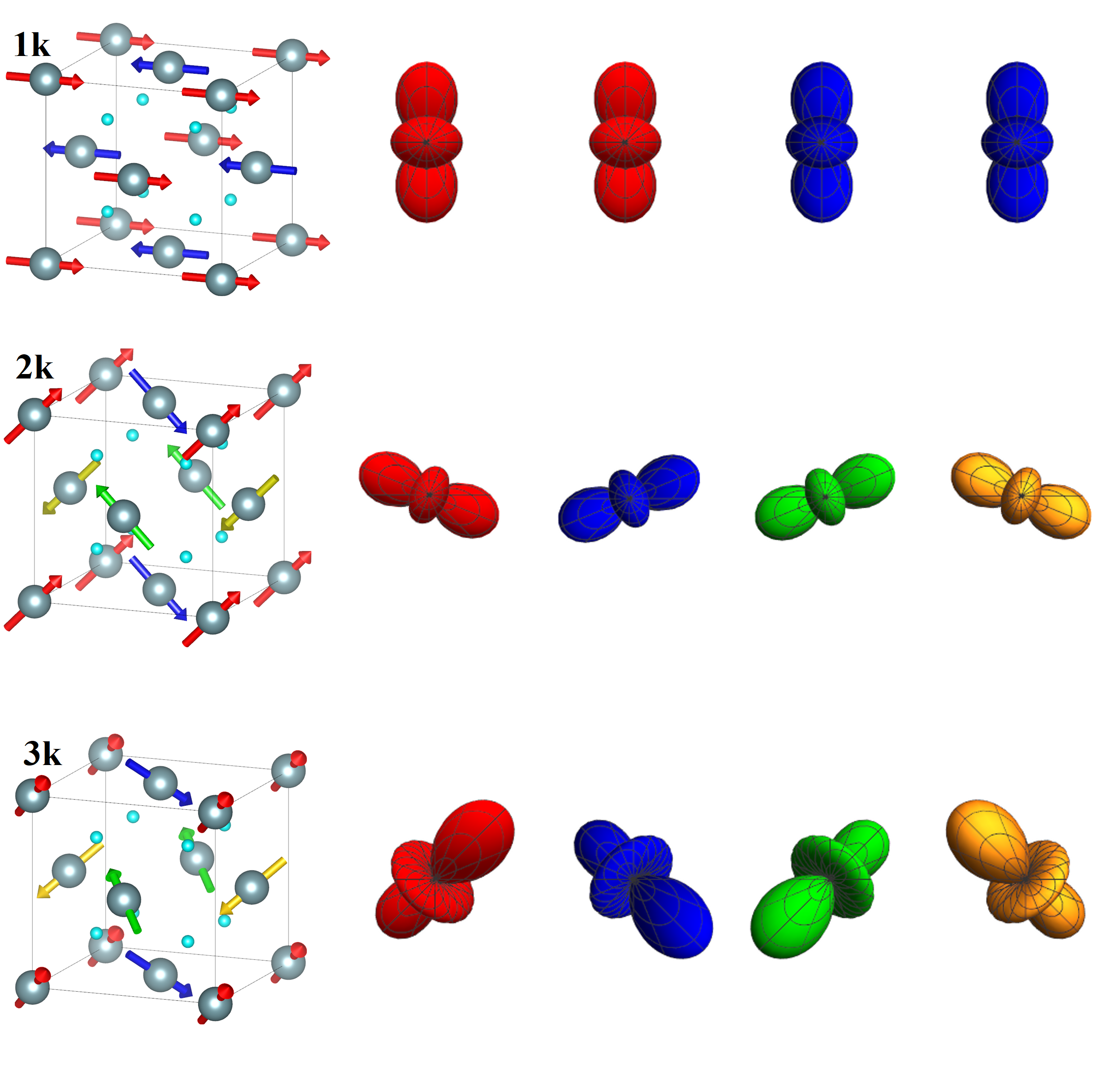} 
		\par\end{centering}
	\caption{1\textbf{k} (upper panel) 2\textbf{k} (middle panel) and 3\textbf{k}
		(lower panel) antiferromagnetic orders in the unit cell of UO$_{2}$.
		Uranium and oxygen sites are shown as large grey and small cyan balls,
		respectively. The quadrupole moments on inequivalent simple-cubic
		sublattices of the magnetic cell obtained by the mean-field solution
		of the {\it ab initio} SE Hamiltonian, eq. (\ref{H_DD}) and (\ref{H_QQ}), at $T=$0 for each antiferromagnetic structure
		are displayed on the right-hand side and colored to indicate the corresponding
		U site in the unit cell.}
	\label{fig:mag_struct} 
\end{figure}

The uranium dioxide is a prototypical example of the MO in actinide
magnetic insulators \cite{Santini2009,Caciuffo2011}. It has a simple
cubic fluorite structure, where U atoms occupy the fcc sublattice
(see Fig.~\ref{fig:mag_struct}). Due to its importance as a nuclear
fuel \cite{Skinner2014} and chemical catalyst \cite{Hutchings1996}
it has been thoroughly studied experimentally. UO\textsubscript{2}
undergoes a first-order phase transition into an antiferromagnetically
(AFM) ordered state at the N\'eel temperature, $T_{N}$ of 30.8~K \cite{Amoretti1989}.
This transition is accompanied by an onset of MO \cite{Wilkins2006}
affecting both phonons and magnons dynamics \cite{Caciuffo2011,Santini2009,Caciuffo1999,Blackburn2005}.
Dynamical Jahn-Teller effects associated with a spin-lattice quadrupolar
coupling is also observed well above $T_{N}$ \cite{Amoretti1989,Caciuffo1999}.

The magnetic structure of UO\textsubscript{2} has been experimental
and theoretical puzzle for a long time. 
The magnetic unit cell of UO$_2$ in the AFM phase contains four inequivalent simple cubic uranium sublattices.
Then
the geometrical frustration of the U fcc sublattice results in three
distinct AFM structures shown in Fig.~\ref{fig:mag_struct} being
degenerate in energy with respect to the usual spin-spin anisotropic
Heisenberg exchange 
\cite{Monachesi1983}. These structures are described, respectively, by
a) the single propagation vector \textbf{k}$=[0,0,1]$ (1\textbf{k}
- collinear structure in the upper panel of Fig.~\ref{fig:mag_struct}),
b) two propagation \textbf{k}-vectors (2\textbf{k}, middle panel
of Fig.~\ref{fig:mag_struct}) with mutually perpendicular orientations
of the magnetic moments in the cubic face plane parallel to the plane
of the \textbf{k}-vectors and, c) three perpendicular \textbf{k}-vectors
(3\textbf{k}, lower panel) with the moments oriented in different
(111) directions \cite{Monachesi1983}.  All three AFM structures have been observed in different cubic uranium monopnictides (UX with X=N,P,As,Sb) \cite{Monachesi1983}. 
The 3\textbf{k} structure has been finally confirmed to be the magnetic
ground state of UO\textsubscript{2} by neutron diffraction and nuclear
magnetic resonance experiments \cite{Burlet1986,Amoretti1989,Ikushima2001}.

The mechanism leading to the stabilization of non-collinear 3\textbf{k}
AFM in UO$_{2}$ has not been clearly identified to date. The crystal field splitting
obtained in various experiments suggests that the ground state of
the U\textsuperscript{4+} ions in UO\textsubscript{2} is a spherically
symmetric $\Gamma_{5}$ triplet well separated from excited CF states
\cite{Amoretti1989,CEFNakkotte} thus the observed AFM structure cannot be due to the single-ion anisotropy. The lattice induced quadrupole-quadrupole
(QQ) coupling might explain the first-order nature of the magnetic
transition in UO\textsubscript{2}, however, it seems to favor 1\textbf{k}-structure
rather than 3\textbf{k} one \cite{Erdos,Solt1980}. Hence, the 3\textbf{k}
AFM should be rather due to a purely electronic mechanism with lattice
distortions subsequently induced by the magnetic ordering \cite{Erdos}.
The electronic quadrupolar superexchange (SE) can in principle stabilize
the 3\textbf{k}-magnetic order in the structurally undistorted high-temperature
phase as suggested by Ref.~ \onlinecite{Mironov}. 
They supported this conjuncture with a rather crude estimation of
SE interactions (SEI) within a semi-empirical kinetic exchange model
. 


A reliable estimation of the QQ superexchange couplings in UO\textsubscript{2}
is thus crucial to unravel the origin of its unusual noncollinear
order. The theoretical evaluation of MIs by {\it ab initio} density-functional-theory
(DFT) methods have a recognized vital importance in the field (see
Refs.~\onlinecite{OpennerJPJ,CompRendusOpeneer} for review). However, the standard
DFT framework in conjunction with local or semi-local exchange correlation
functionals is not applicable to localized U 5$f$ states in UO$_2$.
The DFT+U method, which was extensively employed to study UO$_{2}$
\cite{MagnaniOppeneer,Nordstrom,Zhou2009}, is able to capture this
localization, but only in the symmetry-broken ordered state. Pi \textit{et
	al.\cite{Pi2014}} has recently developed an approach for evaluating
MIs based on a simultaneous flip of multipolar moments on two sites
in a MO state described within DFT+U. Pi \textit{et al.} ~\cite{Pi2014,Pi-PRB}
predicted the spin-wave spectra of UO$_{2}$ in reasonable agreement
with experiment, but their calculated SE QQ interactions are
ferromagnetic and would favor the 1\textbf{k} AFM magnetic order instead
of the 3\textbf{k} one.

Both the high-temperature paramagnetic phase and ordered states of
correlated $f$ compounds can be in principle quantitatively described by combining
DFT with the dynamical mean-field theory (DMFT)\cite{Georges1996}
treatment of localized $f$ shells. This DFT+DMFT method \cite{Anisimov1997,Lichtenstein_LDApp,Kotliar2006}
has been extensively employed to study the electronic structure of
paramagnetic UO$_{2}$ \cite{Kolorench,Lanata2017}. However,
	low symmetries, small energy scales and a vast configurational space
	of MO phases render a direct application of DFT+DMFT to the symmetry-broken phase of UO$_2$ difficult. 
	
In this work we first derive the \textit{ab initio} electronic structure
and CF splitting of UO$_{2}$ in its paramagnetic cubic phase and then apply the linear-response post-processing
of Ref.~\onlinecite{Pourovskii2016} to these converged DFT+DMFT results
evaluating all relevant dipole and multipole SEIs for the CF ground
state. The resulting {\it ab initio} SE Hamiltonian is then solved within the mean-field approximation. We find that its most stable ordered structure is of the non-collinear 3\textbf{k} type and that its stabilization originates from a particular anisotropy of quadrupole-quadrupole SEIs in UO$_2$.

The paper is organized as follows: in the next section we outline the methodology of our electronic structure and superexchange calculations also specifying relevant calculational parameters.
The results of these calculations, namely, the {\it ab initio} superexchange Hamiltonian of UO$_2$ and its mean-field solution, are presented in Sec.~\ref{sec:results}. In Sec.~\ref{sec:discussion} we analyze the calculated superexchange interactions identifying a mechanism for the stabilization of  3\textbf{k} magnetic  order and also study short-range order effects in UO$_2$ above its ordering temperature.

\section{Method}
Our self-consistent in the charge density DFT+DMFT calculations were
carried out employing the approach of Refs.~\onlinecite{Aichhorn2009,Aichhorn2011},
which combines a linearized augmented planewave band structure method
\cite{Wien2k} and the DMFT implementation \cite{Parcollet2015,Aichhorn2016}. The spin-orbit coupling for the UO$_2$ Kohn-Sham band structure was included within the standard second-variation procedure as implemented in Ref.~\onlinecite{Wien2k}, which is expected to be sufficient for the valence (but not semicore) states of uranium. 

Wannier orbitals $\omega_{m\sigma}$ representing U 5$f$ states (where $m$ and $\sigma$ are magnetic and spin quantum numbers, respectively) were constructed from
the manifold of 14 Kohn-Sham 5$f$-like bands located in the vicinity of the Fermi level. The on-site repulsion between these orbitals  was specified by the Slater parameters, F$^0$, F$^2$, F$^4$, and F$^6$. We made use of the standard approximation fixing the ratios of F$^2$/F$^4$ and F$^2$/F$^4$  to the values obtained in Hartree-Fock calculations for the corresponding free ions. We employ the ratios of 1.50 and 2.02, respectively, in good agreement with the values for actinide ions reported, for example, in Ref.~\cite{Moore2009}.  With this choice the values of F$^2$, F$^4$, and F$^6$ are determined by the Hund's rule coupling $J_H$ \cite{Anisimov1997}. We used $F^{0}=$4.5~eV and  $J_{H}=$0.6~eV obtained for UO$_{2}$ in
recent constrained random-phase calculations \cite{Seth2017}.
 SEIs can exhibit a strong sensitivity to the value of $J_{H}$, hence,
to verify the robustness of our results we also performed calculations
with $J_{H}=$ 0.7~eV previously employed in Ref.~\onlinecite{Pi2014}.

The DMFT quantum impurity problem was solved in the quasi-atomic Hubbard-I
approximation (HIA) \cite{hubbard_1}, which is expected to be reasonable
for the paramagnetic high-$T$ phase of the Mott insulator UO$_{2}$.  The hybridization function is neglected within the HIA, and the DMFT impurity problem is reduced to diagonalization of the single-shell Hamiltonian $\hat{H}_{at}=\hat{H}_{1el}+\hat{H}_U=\sum_{mm'\sigma\sigma'}\epsilon_{mm'}^{\sigma\sigma'}f^{\dagger}_{m\sigma} f_{m'\sigma'}+\hat{H}_U$, where $f_{m\sigma}$ ($f^{\dagger}_{m\sigma}$) is the creation (annihilation) operator for the U 5f orbital $m\sigma$, 
 $\hat{H}_U$ is the on-site Coulomb repulsion vertex constructed as described above and $\hat{\epsilon}$ is the non-interacting level position matrix\cite{Lichtenstein_LDApp}. In the DMFT framework $\hat{\epsilon}$  obtained by a high-frequency expansion of the bath Green's function \cite{Pourovskii2007} reads:
\begin{equation}\label{H1el_HI}
	\hat{\epsilon}=-\mu+\langle \hat{H}_{KS} \rangle^{ff} -\Sigma_{\mathrm{DC}}
\end{equation}
where $\mu$ is the chemical potential, $\langle \hat{H}_{KS} \rangle^{ff}=\sum_{\vk \in BZ}\hat{P}_{\vk} H_{KS}^{\vk}\hat{P}_{\vk}^{\dagger}$ is the Kohn-Sham Hamiltonian projected to the basis of 5$f$ Wannier orbitals $\omega_{m\sigma}$ and summed over the Brillouin zone, $\hat{P}_{\vk}$ is the corresponding projector between the KS and Wannier spaces \cite{Aichhorn2009,Aichhorn2016}, $\Sigma_{\mathrm{DC}}$ is the double counting correction term. As the spin-orbit coupling is included in the Kohn-Sham states it naturally appears in $\hat{\epsilon}$ together with the crystal-field splitting. The double-counting correction $\Sigma_{\mathrm{DC}}$ was calculated in the fully-localized
limit \cite{Czyzyk1994} using the atomic occupancy~ \cite{Pourovskii2007}
of the U 5$f^{2}$ shell.  

The DFT+DMFT self-consistent calculations employing the HIA (we abbreviate this framework DFT+HIA below)
were carried out enforcing the uniform occupancy of U 5$f^{2}$ states
within its ground-state multiplet (GSM) in order to suppress the impact
of DFT self-interaction error onto the CF splitting \cite{Delange2017} and at the experimental lattice parameter $a=$5.47 \AA\  of UO$_2$.

In order to evaluate dipole and quadrupole SEIs
acting between U shells in UO$_2$ state we employed 
the method of Ref.~\cite{Pourovskii2016}.
Namely, after having converged DFT+HIA for the symmetry-unbroken paramagnetic state  one evaluates the linear response of the DFT+DMFT grand potential
$\Omega$ to small fluctuations of the on-site density matrix on two
neighboring sites $\vR$ and $\vR'$ with respect to its paramagnetic configuration. These fluctuations are assumed to be limited to the ground-state (GS) crystal-field (CF) level for the cases when the magnitude of SEIs is much smaller than that of the CF splitting. 

The corresponding variational derivative of the DFT+DMFT grand potential with respect to such fluctuations    $\frac{\delta^{2}\Omega}{\delta\rho^{\alpha\beta}({\bf R})\delta\rho^{\gamma\delta}({\bf R'})}$ is then identified as the matrix element $\langle\alpha\gamma|V(\vR'-\vR)|\beta\delta\rangle$ of  SEI $V(\vR'-\vR)$ between the two-site states $|\alpha\gamma\rangle$ and $|\beta\delta\rangle$. The first and second letter in $|...\rangle$ labels  a given CF state of the CF GS level on the ion $\vR$ and $\vR'$, respectively.
The lowercase Greek letters designate states within the GS CF level, $\hat{\rho}$ is the density matrix for the GS CF level.  The dependence of $V$ on $\vR'-\vR$ only is due to the translational
invariance.  As shown in Ref.~\cite{Pourovskii2016},
$\frac{\delta^{2}\Omega}{\delta\rho^{\alpha\beta}({\bf R})\delta\rho^{\gamma\delta}({\bf R'})}=\frac{1}{\beta}Tr\left[G_{{\bf RR'}}\frac{\delta\Sigma}{\delta\rho^{\gamma\delta}}G_{{\bf R'R}}\frac{\delta\Sigma}{\delta\rho^{\alpha\beta}}\right]$,
where the variational derivative of the local self-energy $\Sigma$
with respect to a given fluctuation $\rho^{\alpha\beta}$ of the density matrix is evaluated
analytically within the HIA. The inter-site Green's
function (GF) $G_{{\bf RR'}}$ is obtained by a Fourier transform
of the lattice GF projected to the basis of correlated 5$f$ orbitals. 

\section{Results}\label{sec:results}

\begin{figure}[!b]
	\begin{centering}
		\includegraphics[width=0.9\columnwidth]{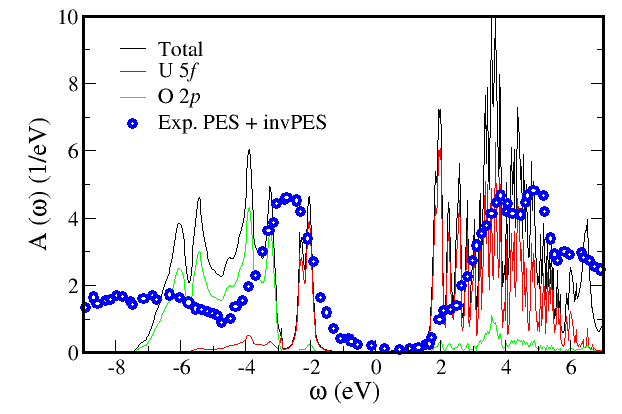} 
		\par\end{centering}
	\caption{The DFT+DMFT spectral function of UO$_2$ within the Hubbard-I approximation. The black, red, and green lines are the total, partial U 5$f$ and O 2$p$ spectral functions, respectively. The experimental emission and  bremsstrahlung isochromat spectra of Ref.~\cite{Yu2011} are displayed by blue circles.}
	\label{UO2_spectra} 
\end{figure}

We start by discussing the electronic structure and many-electron states of U 5$f$ shell as obtained by the DFT+HIA method for the paramagnetic phase of UO$_2$.
In Fig.~\ref{UO2_spectra} we display the calculated valence-band spectral function compared to  recent photoemission (PES) and  bremsstrahlung isochromat spectra (BIS) of Ref.~\onlinecite{Yu2011}. These experimental measurements  employed high photon energies thus enhancing the relative spectral weight of 5$f$ features.  Our calculated valence-band spectral function is in an overall qualitative agreement with the experimental results of  Ref.~\onlinecite{Yu2011}. The width of U 5$f$ upper Hubbard band is mostly due to multiplet effects and in agreement with the experimental spectra, while the width of lower Hubbard band is due to hybridization effects and underestimated due to well-known limitations of the Hubbard-I approximation \cite{Dai2005}. The overall splitting between upper and lower Hubbard bands and the multiplet splitting of excited states that are crucial to correctly capture the superexchange phenomenon are quantitatively well reproduced by our approach.

As outlined in the Method section, the DMFT impurity problem within the HIA is reduced to a single 5$f$ shell Hamiltonian $\hat{H}_{at}$, its one-electron level positions (\ref{H1el_HI})  includes both the spin-orbit and crystal-field effects. The value of spin-orbit coupling parameter $\lambda=$0.235~eV extracted from our converged $\hat{\epsilon}$ is in agreement with Hartree-Fock calculations for free U ion \cite{Ogasawara1991}. It is expected that $\lambda$ being an essentially intra-atomic quantity is not significantly affected by the solid-state environment. By diagonalizing $\hat{H}_{at}$ 
we obtained the $^{3}H_{4}$ ground-state multiplet (GSM) of U 5$f^{2}$
shell with the $\Gamma_{5}$ triplet being the CF ground state; the
exited doublet $\Gamma_{3}$, triplet $\Gamma_{4}$, and singlet $\Gamma_{1}$
predicted to be 193, 197, and 207 meV higher in energy, respectively.
Our theoretical CF splitting is thus in good agreement with experimental
measurements \cite{CEFNakkotte} that found the splitting of 150 to
180 meV meV between the $\Gamma_{5}$ ground state and densely-spaced
exited CF levels, as well as with previous DMFT calculations of Ref.~\cite{Kolorench}.
This CF splitting is much higher than $T_{N}$ of UO$_{2}$, hence,
the impact of exited multiplets on the magnetic order can be neglected. 

The calculated $\Gamma_{5}$ eigenstates in the $|J;m_{J}\rangle$
basis 
\begin{flalign}
|1\rangle= & 0.908|4;+3\rangle-0.343|4;-1\rangle-0.032|5;-5\rangle\label{Gamma5_WF}\\
|0\rangle= & 0.686|4;+2\rangle-0.686|4;-2\rangle-0.033|5;-2\rangle\nonumber \\
& -0.033|5;+2\rangle\nonumber \\
|-1\rangle= & -0.908|4;-3\rangle+0.343|4;+1\rangle-0.032|5;+5\rangle\nonumber 
\end{flalign}
feature a small admixture of high-energy multiplets. 


We calculated the SEIs between the $\Gamma_5$ states (\ref{Gamma5_WF}) by the approach of Ref.~\cite{Pourovskii2016} outlined in the method section.
There are in total $3^{4}=81$ SEIs
$\langle\alpha\gamma|V(\vR'-\vR)|\beta\delta\rangle$ for each U-U
bond. We have subsequently transformed these interactions to more
conventional SE couplings between the spherical tensor dipole and
quarupole moments. The $\Gamma_{5}$ triplet (effective angular momentum  $\tilde{J}=1$) can support both dipole
and quadrupole moments\footnote{The observable dipole magnetic and quadrupole moments are related
	to the corresponding tensor moments defined in the $\Gamma_{5}$ basis
	(\ref{Gamma5_WF}) by the prefactors 2.88, 0.153 and 0.217 for the
	dipole, quadrupole $e_{g}$ and quadrupole $t_{2g}$, respectively.}. The SEIs between those moments were obtained using the transformation $\sum_{\alpha\beta\gamma\delta}\langle\beta\delta|V(\vR'-\vR)|\alpha\gamma\rangle O_{\alpha\beta}^{LM}O_{\gamma\delta}^{L'M'}=V_{MM'}^{LL'}(\vR'-\vR)$,
where $O_{\alpha\beta}^{LM}$ is the $\alpha\beta$ matrix element
of the real spherical tensor for the  effective angular momentum $\tilde{J}=1$ \cite{Santini2009,Pourovskii2016}
of the rank $L=1$ (dipole) or 2 (quadrupole) and projection $M$.
$V_{MM'}^{LL'}(\vR'-\vR)$ is the resulting SEI between the multipoles
$LM$ and $L'M'$ located at the sites $\vR$ and $\vR'$, respectively.

Thus calculated SE Hamiltonian for the nearest-neighbor (NN) U-U bond
$\vR'-\vR=[1/2,1/2,0]$ is of the form $H_{SE}=H_{DD}+H_{QQ}$, where
the dipole-dipole (DD) and QQ contributions (in the global coordinate
system) read 
\begin{flalign}
H_{DD} & =V\sum_{M=x,y}\hat{O}_{\vR}^{M}\hat{O}_{\vR'}^{M}+V'\hat{O}_{\vR}^{z}\hat{O}_{\vR'}^{z}\label{H_DD}\\
& +V_{x,y}[\hat{O}_{\vR}^{x}\hat{O}_{\vR'}^{y}+\hat{O}_{\vR}^{y}\hat{O}_{\vR'}^{x}],\nonumber 
\end{flalign}
\beq\label{H_QQ}
H_{QQ}=\sum_{M \in t_{2g},e_g}V^q_{M}\hat{O}_{\vR}^{M}\hat{O}_{\vR'}^{M}+V^q_{xz,yz}[\hat{O}_{\vR}^{xz}\hat{O}_{\vR'}^{yz}+\hat{O}_{\vR}^{yz}\hat{O}_{\vR'}^{xz}].
\eeq The number of independent SE couplings is seen to be significantly
reduced due to the cubic symmetry of the problem. Hence, for brevity
we omit the rank $L$ in the real tensors, as the projection $M$
is sufficient to identify them unambiguously, and suppress superfluous
indices for $V$. The QQ SEI are labeled by the superscript $q$.
SE Hamilonians for other NN bonds are easily obtained from (\ref{H_DD})
and (\ref{H_QQ}) by symmetry. Our choice for the spherical tensors representing the dipole and quadrupole DOFs of the $\Gamma_5$ triplet is in agreement with Refs.~\cite{Pi2014,Pi-PRB}, however, following Santini {\it et al.}~\cite{Santini2009}we employ $\hat{O}$ to denote real spherical tensors instead of $\hat{T}$ in Refs.~\cite{Pi2014,Pi-PRB}. The operator notation is also employed in the literature \cite{Mironov,Carretta2010,Caciuffo2011} for the low-energy Hamiltonian of UO$_2$. SEIs in the operator formalism are related to the tensor SEIs in (\ref{H_QQ}) by a simple renormalization that we specify in Appendix~\ref{appen:tensors_to_ops}.  The interactions of next-nearest neighbors
(NNN) are an order of magnitude smaller and induce no qualitative
changes, they are listed in Appendix~\ref{appen:NNN}. More
distant SEIs are negligible. The calculated NN SEIs for two values
of $J_{H}$ are listed in Table~\ref{Tab_V}. One may see that the
variation in $J_{H}$ has a rather insignificant impact on the SEIs.
Unless explicitly mentioned otherwise, we use the SEIs for $J_{H}=$0.6~eV
in all calculations below.

  \begin{table}
  	\centering
  	\caption{\label{Tab_V} Calculated U-U nearest-neighbor interactions for the [1/2,1/2,0] bond (meV) as a function of the Hund's rule coupling $J_H$ .}
  	\begin{tabular}{c|| c c c || c c c c c }
  		$J_H$ (eV)       & $V$ & $V'$ & $V_{x,y}$ & $V^q_{xy}$ & $V^q_{xz(yz)}$ &  $V^q_{x^2-y^2}$ & $V^q_{z^2}$  & $V^q_{xz,yz}$ \\ \hline
  		0.6   & 1.42 &  3.85 & -0.67 & 0.18 & 0.01 & -0.16 & 0.14 & 0.04 \\
  		0.7  & 1.39 &  3.73 & -0.69 & 0.20 & 0.01 &  -0.18 & 0.17 & 0.04 \\
  	\end{tabular}
  \end{table}

\begin{figure}[!tb]
	\begin{centering}
		\includegraphics[width=0.97\columnwidth]{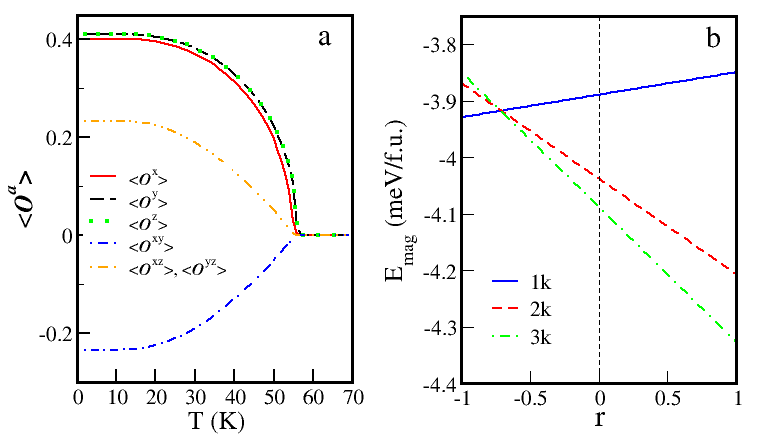} 
		\par\end{centering}
	\caption{a. The expectation values of dipole and $t_{2g}$ quadrupole tensors
		as a function of temperature. A phase transition at $T=$56~K is
		clearly seen. b. The mean-field magnetic energy $E_{mag}$ at zero
		temperature as a function of the anisotropy parameter $r$ of the
		QQ SE, see text. }
	\label{fig_ordered_state} 
\end{figure}

We have subsequently solved the calculated \textit{ab initio} SE Hamiltonian
including NN and NNN coulings within the mean-field approximation
(MFA) implemented in Ref.~\onlinecite{Rotter2004}. We considered three
structures shown in Fig.~\ref{fig:mag_struct} as well as all single-\vk~magnetic
structures realizable within the 4$\times$4$\times$4 fcc supercell.
A clear phase transition is observed in the evolution of specific
heat at about $T_{N}=$56~K (with only NN SEIs $T_N=60$~K)  accompanied by appearance of a non-zero
on-site dipole moment oriented along the $\langle111\rangle$ direction
and quadrupole moments of the $t_{2g}$ irreducible representation
(IREP) as shown in Fig.~\ref{fig_ordered_state}a \footnote{Using the SEIs evaluated with $J_{H}=0.7$~eV we obtain the same
	ordered structures at almost identical temperature of 54~K.}. The obtained magnetic and quadrupole orders correspond to the 3\vk-structure
plotted the lower panel of Fig.~\ref{fig:mag_struct}, which is the
experimental ordered structure of UO$_{2}$. Predicted $T_{N}$ is
substantially higher than the experimental first-order transition
temperature of 30.8~K. A large overestimation of $T_{N}$ in the
MFA is expected for the fcc lattice due to its geometric frustration\cite{Minor1988,Diep1989}.

\section{Analysis and discussion}\label{sec:discussion}

Let us now analyze the calculated SEIs in order to identify the origin
of 3\vk-structure stabilization with respect to the competing 1\vk~and
2\vk~ones. The DD interactions are antiferromagnetic and very asymmetric.
With $J'<J<0$ all three AFM structures shown in Fig.~\ref{fig:mag_struct}
become degenerated with respect to $H_{DD}$ (\ref{H_DD})
having the same ordering energy $E_{mag}=-V'=-3.85$ meV/f.u. in the
mean-field approximation\cite{Erdos,BakJensen}. The quarupole orders shown in the rhs of
Fig.~\ref{fig:mag_struct} are obtained by solving the full SE NN
Hamiltonian for the AFM state of a given type.
With the calculated SE QQ interactions from Table~\ref{Tab_V} the
QQ contribution to the ground state energies is equal to 0.010, -0.047,
and -0.060 meV/(f.u.) for the 1\vk, 2\vk, and 3\vk~orders, respectively.

Therefore, we conclude that the QQ SEI are stabilizing the experimentally observed non-collinear 3\textbf{k} magnetic structure in the absence of SL mediated contribution; this order of a purely electronic origin and would subsequently results in the Jahn-Teller distortion. The SL QQ coupling, however, might be essential for the full description of the relative stability of ordered magnetic structures as well as for the spin dynamics in UO$_2$ \cite{Carretta2010,Caciuffo2011}. 

\begin{table}[b]
	\caption{\label{tab_comp}Comparison of the SEI calculated in the present work
		with previous DFT+U calculations of Ref.~\cite{Pi-PRB}, and the
		values of Ref.~\cite{Caciuffo2011} from a fit of the experimental
		spin-wave spectra . Following Refs. \cite{Caciuffo2011,Pi-PRB} we
		define the isotropic part of DD and QQ SEIs as $V'$ and $V_{xy}^{q}$,
		respectively. and the corresponding dimensionless anisotropy parameters
		$\delta^{d/q}$ defined as $\delta^{d}=V/V'$ and $\delta^{q}=V_{yz}^{q}/V_{xy}^{q}$,
		respectively. Refs.\cite{Caciuffo2011,Pi-PRB} estimated only the
		SEI relevant for the 3\textbf{k}-structure, thus only those four parameters
		are available for comparison (Note that Ref.~\cite{Caciuffo2011}
		assumed $\delta^{q}=\delta^{d}$).}
	\begin{ruledtabular} %
		\begin{tabular}{lcrrr}
			& $V'$  & \multicolumn{1}{c}{$\delta^{d}$} & $V_{xy}^{q}$  & $\delta^{q}$\tabularnewline
			This work  & 3.85  & 0.37  & 0.18  & 0.22\tabularnewline
			Ref.\cite{Caciuffo2011}  & 3.1  & 0.25  & 1.9  & 0.25\tabularnewline
			Ref.\cite{Pi-PRB}  & 1.70  & 0.3  & -3.10  & 0.9\tabularnewline
		\end{tabular}\end{ruledtabular} 
	\end{table}
	
	In Table~\ref{tab_comp} we compare our results to previous theoretical
	and experimental estimates of SEIs in UO$_{2}$. Our DD SE is very
	close to the fit of experimental spin-waves spectra of Ref.~\cite{Caciuffo2011},
	however, our QQ SEIs are much smaller. The qualitative difference
	with the DFT+U results \cite{Pi-PRB} is in the sign of the QQ interactions.
	The negative sign predicted in Ref.~\cite{Pi-PRB} would stabilize
	1k-order having NN $\langle\hat{O}_{\vR}^{M}\hat{O}_{\vR'}^{M}\rangle=0$
	for all $M$ belonging to the $t_{2g}$ IREP ($xy$, $xz$, $yz$).
	AF $t_{2g}$ SEIs favor the 3\textbf{k}-structure because of a larger
	angle between ordered quadrupoles in this case as compared to the
	2\textbf{k}-structure, where one third of NN pairs have parallel quadrupole
	moments and the 1\textbf{k}-structure, where all quadrupole moments
	are parallel (see Fig.~\ref{fig:mag_struct}).
	
	
	The magnitude of SEIs acting between the $e_{g}$ quadrupoles has
	not been evaluated in Ref.~\cite{Pi-PRB} neither can it be estimated
	from the spin-wave dispersion, as $\langle\hat{O}_{\vR}^{M}\hat{O}_{\vR'}^{M}\rangle=0$
	for $M=z^{2}$ and $x^{2}-y^{2}$ in the experimental 3\textbf{k}
	AFM structure. However, the contribution of $e_{g}$ SEIs is non-zero
	for the 1\textbf{k} and 2\textbf{k} competing orders 
	thus impacting the relative stability of magnetic structures. 
	
	In order to further clarify the impact of QQ SEIs on the relative stability of these three structures one may evaluate their single-site mean-field Hamiltonian in a local coordinate frame\cite{Carretta2010}, in which the on-site dipole moment is parallel  to the local $z$ axis. In such a frame 
	 only the $z^2$ quadrupole is active, while other local quadrupole moments are zero thus simplifying the analysis (see Appendix~\ref{appen:tensors_to_ops} for the definition of quadrupole moments in terms of spin operators). The mean-field Hamiltonian reads $\hat{H}_{MF}=\hat{H}_{DD}^{MF}+\hat{H}_{QQ}^{MF}$, where the dipole-dipole contribution $\hat{H}_{DD}=-4V'\langle \hat{O}_l^{z}\rangle  \hat{O}_l^{z}$ is the same for all  three structures. $ \hat{O}_l^{z}$ is the $z$ projection of the dipole tensor operator in the local frame, its expectation value $\langle \hat{O}_l^{z}\rangle=1/\sqrt{2}$ at the full saturation.
	
	 The QQ term $\hat{H}_{QQ}^{MF}$ reads $V_{QQ} \langle \hat{O}_l^{z^2}\rangle  \hat{O}_l^{z^2}$, where the quadrupole operator $\hat{O}_l^{z^2}$ is defined in the local frame in the same way as the dipole one. The mean-field QQ coupling $V_{QQ}$ is equal to  $6V^q_{e_g}$,  $(\frac{3}{8\sqrt{2}}+\frac{9}{8})V^q_{e_g}-3V^q_{xy}$, and $-4V^q_{xy}$ for the 1\vk, 2\vk, and 3\vk~orders, respectively. Here we designate as $V^q_{e_g}=V^q_{z^2}+V^q_{x^2-y^2}$  the summed diagonal SEI between $e_g$ quadrupoles . One sees that the relative stability of the structures is determined by the relative magnitudes of $V^q_{e_g}$ and the in-plain coupling between $t_{2g}$ quadrupoles $V^q_{xy}$. In particular, the opposite signs of our calculated  $V^q_{z^2}$ and $V^q_{x^2-y^2}$ result in the magnitude of 
	 $V^q_{e_g} =$-0.02~meV that is much smaller than $V^q_{xy}=$0.18~meV. While $V^q_{e_g}$ is negative and does help stabilizing 1\textbf{k} and  2\textbf{k}  structures, its contribution is overweighted by a larger prefactor for $V^q_{xy}$ in the case of 3\vk.  	Therefore, it is the particular
	 anisotropy of QQ SEIs of UO$_{2}$ with a larger magnitude of positive
	 $V_{t_{2g}}^{q}$ that is at the origin of 3\textbf{k}-order
	 in UO$_{2}$.
	
	From the mean-field Hamiltonian derived above one may easily evaluate the effect of a variation in the relative value of the QQ SEIs $V^q_{e_g}$ and $V^q_{z^2}$ on the ground-state magnetic structure of UO$_2$. One may introduce renormalized 
	SEIs $V_{e_g}^{q}(1-r)$ and $V^q_{xy}(1+r)$
	with $r\in[-1:1]$. Hence, $r=0$ corresponds to the actual calculated
	QQ SEIs, while at $r=-1$ ($1$) only $V_{e_g}^{q}$
	($V^q_{xy}$) are non-zero. The resulting evolution of the mean-field ordering energy $E_{mag}$ vs. $r$ is plotted in Fig.~\ref{fig_ordered_state} b. One finds that  the 1\textbf{k}
	is stabilized with $r\to-1$, while the actual 3\textbf{k} is stabilized
	in the opposite limit. At $r\approx-0.713$ one obtains a transition between the 1\textbf{k} and 3\textbf{k} orders.  Interestingly, the 2\textbf{k}  structure is unstable relative to  the 1\textbf{k} order for  $r < -0.707$ and  relative to the 3\textbf{k} one for $r>-0.732$, meaning that over the whole range of $r$ the ground state is never of the 2\textbf{k} type.

	\begin{figure}[!tb]
		\begin{centering}
			\includegraphics[width=0.9\columnwidth]{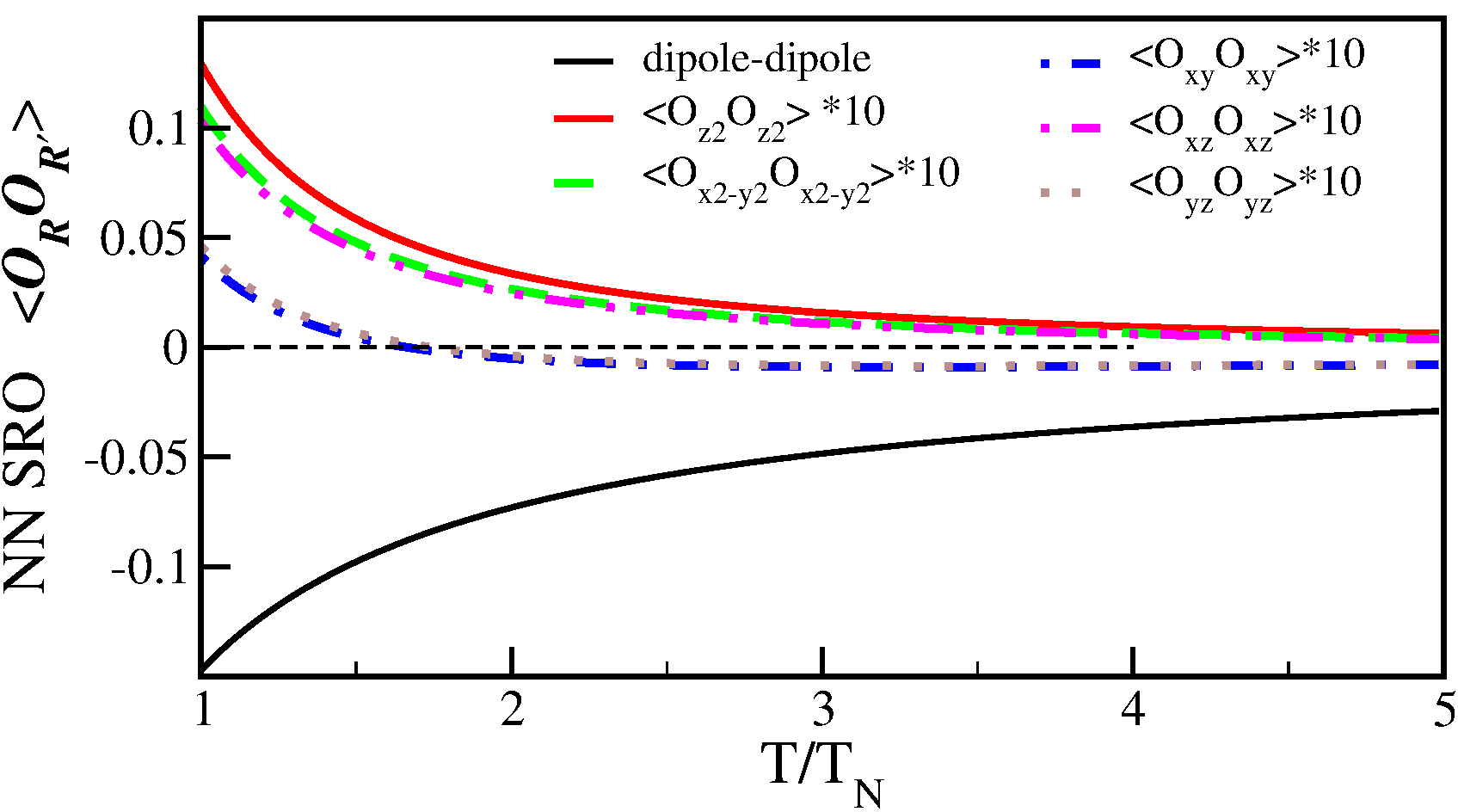} 
			\par\end{centering}
		\caption{The dipole-dipole and quadrupole-quadrupole nearest-neighbor pair
			correlation functions above $T_{N}$. The quadrupole-quadrupole pair
			correlation functions are multiplied by 10.}
		\label{fig_SRO} 
	\end{figure}
	
	The phase transition in UO\textsubscript{2} is of the first order
	and dynamical Jahn-Teller effects are also observed well above $T_{N}$
	\cite{Caciuffo1999,Amoretti1989} hinting at a non-negligible short-range
	order (SRO) present in UO$_{2}$. We have analyzed SRO effects above
	the N\'eel temperature using an Oguchi-like method \cite{Oguchi}. To
	this end we diagonalized the \textit{ab initio} SE Hamiltonian, eqs.
	(\ref{H_DD}) and (\ref{H_QQ}), with the SEIs from Table~\ref{Tab_V},
	for each NN pair of U ions. We then calculated the DD and QQ pair
	correlation functions $\langle\hat{O}_{\vR}^{M}\hat{O}_{\vR'}^{M'}\rangle$
	by averaging them over all NN bonds. The calculated NN pair correlation
	functions vs. $T/T_{N}$ are shown in Fig.~\ref{fig_SRO}. Strong
	dipole SRO effects are clearly observed well above N\'eel temperature as
	expected for the frustrated fcc lattice \cite{Khmelevskyi2012}.  The dominating  AFM dipole SRO forces a ferroquadrupole SRO for both the $t_{2g}$ and $e_g$ quadrupoles for $T > T_N$ as one sees in  Fig.~\ref{fig_SRO}. The constrain of anti-parallel orientation of the neighboring dipole moments is lifted in the ordered state by the AFM frustration. The $t_{2g}$ quarupole order is then antiferro due to the corresponding sign of QQ SEIs,  while the $e_g$ pair correlation functions are zero. Hence, the structure of QQ pair correlation
	function below and above the phase transition is qualitatively different. 
	This observation has two important consequences. First,  a SRO that is opposite to the corresponding pair correlation function in the ordered state is associated with a first-order magnetic phase transition  \cite{Nagaev1979,Kovalenko1982,Nagaev1982}. This hints at a purely electronic SE mechanism for  the observed first-order type of magnetic transition in UO$_2$. Second, the dynamical Jahn-Teller distortions above $T_{N}$ might be quite
	different from the static one in the AFM phase. The last prediction can be possibly verified in future experimental studies.
	
	\section{Conclusions}
	
	In conclusion, our calculations
	point out at the anisotropy of quadrupole superexchange as a likely origin
	of non-collinear 3\textbf{k} antiferromagnetic order in UO$_{2}$ and the first-order type of the corresponding N\'eel transition.
	The present \textit{ab initio} approach seems to be highly promising
	for studies of other localized $f$-electron systems featuring complex
	unexplained magnetic or \char`\"{}hidden\char`\"{} orders and local
	multipole degrees of freedom.
	
	\section{Acknowledgments} LP acknowledges the support of the
	European Research Council Grant No. ERC-319286-QMAC and computational resources provided by the Swedish National Infrastructure for Computing (SNIC). S.K. is grateful to Ecole Polytechnique and the Austrian CMS for financial support and to CPHT for its hospitality.

\appendix

\section{Superexchange in spherical-tensor and angular-moment-operator formalisms}\label{appen:tensors_to_ops}

In the present work we represent the dipole and quadrupole DOF of the $\Gamma_5$ triplet by the real-valued spherical tensors\cite{Santini2009,Blum_DM}  $\hat{O}$  similarly to Refs.~\onlinecite{Pi2014,Pi-PRB}. Some authors \cite{Mironov,Carretta2010,Caciuffo2011} employ instead the conventional angular-moment and quadrupole operators. The SEIs defined in the two formalisms are related to each other by simple prefactors. Namely, for the effective angular momentum $\tilde{J}=1$ the dipole spherical tensors $\hat{O}^{M}=\hat{S}_M/\sqrt{2}$, where $\hat{S}_M$ is the angular-moment operator for the same projection $M=x$, $y$, or $z$, see Ref.~\onlinecite{Blum_DM}. Hence, one sees that the dipole-dipole SEIs in the angular-momentum formalism $J_{MM'}$ are related to our spherical-tensor ones by the prefactor 1/2, $J_{MM'}\equiv V_{MM'}/2$.  

The quadrupole spherical tensors for  $\tilde{J}=1$ can be expressed as products of dipole ones:
\begin{equation}
\begin{aligned}[l]
\hat{O}^{xy}=-\sqrt{2}\left(\hat{O}^x\hat{O}^y+\hat{O}^y\hat{O}^x\right) \\
\hat{O}^{xz}=\sqrt{2}\left(\hat{O}^x\hat{O}^z+\hat{O}^z\hat{O}^x\right) \\
\hat{O}^{yz}=\sqrt{2}\left(\hat{O}^y\hat{O}^z+\hat{O}^z\hat{O}^x\right) \\
\hat{O}^{z^2}=\sqrt{2/3}\left(3\left(\hat{O}^z\right)^2-1\right) \\
\hat{O}^{x^2-y^2}=\sqrt{2}\left(\hat{O}^x\hat{O}^x-\hat{O}^y\hat{O}^y\right).
\end{aligned}
\end{equation}
They are converted to the corresponding quadrupole operators employed by Refs.~\onlinecite{Mironov,Carretta2010,Caciuffo2011} by multiplying them by $\sqrt{6}$ for $z^2$ and $\sqrt{2}$ for all other projections. Hence, the corresponding conversion factors between the spherical-tensors QQ SEIs $V^q_M$ (Tables~\ref{Tab_V} and \ref{Tab_V_NNN}) and QQ interactions $K_M$ in Refs.~\onlinecite{Mironov,Carretta2010,Caciuffo2011} are $1/6$ for the SEI coupling $M=z^2$, $K_{z^2}\equiv V_{z^2}/6$, and $1/2$ for all other $V^q_M$.

The ordered states analyzed in this paper are specified by the following expectation values of the dipole tensors:

1\textbf{k} structure:
\begin{equation}
\langle \hat{O}_x^{\vR} \rangle=\frac{e^{i2\pi R_y}}{\sqrt{2}}; \langle \hat{O}_y^{\vR} \rangle=0;  \langle \hat{O}_z^{\vR} \rangle=0,
\end{equation}

2\textbf{k} structure:
\begin{equation}
\langle \hat{O}_x^{\vR} \rangle=\frac{e^{i2\pi R_z}}{2}; \langle \hat{O}_y^{\vR} \rangle=0;  \langle \hat{O}_z^{\vR} \rangle=\frac{e^{i2\pi R_y}}{2},
\end{equation}

3\textbf{k} structure:
\begin{equation}
\langle \hat{O}_x^{\vR} \rangle=\frac{e^{i2\pi R_z}}{\sqrt{6}}; \langle \hat{O}_y^{\vR} \rangle=\frac{e^{i2\pi R_x}}{\sqrt{6}};  \langle \hat{O}_z^{\vR} \rangle=\frac{e^{i2\pi R_y}}{\sqrt{6}},
\end{equation}
where $\vR$ is the lattice vector in units of the lattice parameter $a$.

\section{Next-nearest-neighbors superexchange interactions in UO$_2$}\label{appen:NNN}

The calculated SE Hamiltonian for the next-nearest-neighbor bond [001] $H^{NNN}_{SE}=H^{NNN}_{DD}+H^{NNN}_{QQ}$ reads
\beq
H^{NNN}_{DD} =V\sum_{M=x,y}\hat{O}_{\vR}^{M}\hat{O}_{\vR'}^{M}+V'\hat{O}_{\vR}^z\hat{O}_{\vR'}^z
\eeq
\beq
H^{NNN}_{QQ}=\sum_{M \in t_{2g},e_g}V^q_{M}\hat{O}_{\vR}^{M}\hat{O}_{\vR'}^{M}
\eeq
where the dipole-dipole (DD) and QQ contributions take a simpler form compared to the nearest-neighbor SE  Hamiltonian  (eqs. 2 and 3 of the main text)   due to the absence of off-diagonal terms. The SE Hamiltonains for other NNN bonds are obtained from that for [001] by the corresponding rotations, that amounts  in the case of $H^{NNN}_{DD}$ to permutations of the $x$, $y$ and $z$ labels.  The   $L=2$ tensors in $H^{NNN}_{QQ}$ transform upon these rotations like the corresponding  $l=2$ real spherical harmonics.

The calculated values of the NNN SEIs are listed in Table~\ref{Tab_V_NNN}.  By comparing it with Table~I of the main text one sees that the NNN SEIs are about one order of magnitude smaller compared to the NN ones.

\begin{table}[h]
	\centering
	\caption{\label{Tab_V_NNN} Calculated U-U next-nearest-neighbor interactions for the [0,0,1] bond (meV) for $J_H=0.6$~eV .}
	\begin{tabular}{ c c || c c c c }
		$V$ & $V'$  & $V^q_{xy}$ & $V^q_{xz(yz)}$ &  $V^q_{x^2-y^2}$ & $V^q_{z^2}$   \\ \hline
		0.143 &  0.156  & 0.004 & -0.015 & -0.003 & 0.053 \\
	\end{tabular}
\end{table}


\end{document}